\documentclass[aps,twocolumn,showpacs,prb,superscriptaddress]{revtex4-2}

\usepackage{amsmath}   
\usepackage{amssymb}
\usepackage{braket}
\usepackage{leftidx}
\usepackage{graphicx}    
\usepackage{verbatim}   
\usepackage{color}         
\usepackage{subfigure}  
\usepackage{dcolumn}    
\usepackage{textcomp}
\usepackage{lipsum}
\usepackage{multirow}
\usepackage{siunitx}
\usepackage{tabularx}
\usepackage{graphicx,amsmath,amssymb,bm}
\usepackage{hyperref,verbatim}
\usepackage{latexsym}
\usepackage{bbold}
\usepackage{siunitx}

\renewcommand{\eqref}[1]{Eq.~(\ref{#1})}

\RequirePackage[usenames,dvipsnames]{xcolor}
\usepackage{soul}

\begin{document}

\title{Unveiling temperature dependence mechanisms of perpendicular magnetic anisotropy at Fe/MgO interfaces}

\author{Fatima Ibrahim}
\email{fatima.ali.ibrahim@hotmail.com}  
\affiliation{Univ. Grenoble Alpes, CEA, CNRS, SPINTEC, 38000 Grenoble, France}
\author{Ali Hallal} 
\affiliation{Univ. Grenoble Alpes, CEA, CNRS, SPINTEC, 38000 Grenoble, France}
\author{Alan Kalitsov} 
\affiliation{Western Digital Technologies, Inc., 5601 Great Oaks Parkway, San Jose, CA 95119, USA}
\author{Derek Stewart}
\affiliation{Western Digital Technologies, Inc., 5601 Great Oaks Parkway, San Jose, CA 95119, USA}
\author{Bernard Dieny} 
\affiliation{Univ. Grenoble Alpes, CEA, CNRS, SPINTEC, 38000 Grenoble, France}
\author{Mairbek Chshiev}  
\email{mair.chshiev@cea.fr}
\affiliation{Univ. Grenoble Alpes, CEA, CNRS, SPINTEC, 38000 Grenoble, France}
\affiliation{Institut Universitaire de France (IUF), 75231 Paris, France}

\begin{abstract}
The perpendicular magnetic anisotropy at magnetic transition metal/oxide interfaces is a key element in building out-of-plane magnetized magnetic tunnel junctions for spin-transfer-torque magnetic random access memory (STT-MRAM). Size downscaling renders magnetic properties more sensitive to thermal effects. Thus, understanding the temperature dependence of the magnetic anisotropy becomes crucial. In this work, we theoretically address the correlation between temperature dependence of magnetic anisotropy and magnetization in typical Fe/MgO-based structures. In particular, the possible mechanisms behind the experimentally-reported deviations from the Callen and Callen scaling power law are analyzed. At ideal interfaces, first-principles calculations reveal (i) small high-order anisotropy constants compared to first order and (ii) enhanced exchange constants. Considering these two intrinsic effects in the atomistic simulations, the temperature-dependence of the total and layer-resolved anisotropy are found to follow the Callen and Callen scaling power law, thus ruling out an intrinsic microscopic mechanism underlying deviations from this law. Besides, two possible extrinsic macroscopic mechanisms are unveiled namely the influence of the dead layer, often present in the storage layer of STT-MRAM cells, and the spatial inhomogeneities of the interfacial magnetic anisotropy. About the first mechanism, we show that the presence of a dead layer tends to reduce the scaling exponents. In the second mechanism, increasing the percentage of inhomogeneity in the interfacial perpendicular magnetic anisotropy is revealed to decrease the scaling exponent. These results allow us to coherently explain the difference in scaling exponents relating anisotropy and magnetization thermal variations reported in earlier experiments. This is crucial for the understanding of the thermal stability of the storage layer magnetization in STT-MRAM applications.
\end{abstract}


\maketitle


\section{Introduction}

The evolution in the magnetic random access memory (MRAM) technologies was made possible thanks to the  fundamental research breakthroughs in spintronic phenomena and materials development. Intensive efforts focus on spin transfer torque (STT)~[\cite{Slonczewski96,Berger96}] that enables current-induced switching and thus better downscalability compared to field-written MRAM~[\cite{Kawahara12,Kent15,Apalkov16,Khvalkovskiy13}]. More precisely, perpendicular STT-MRAMs where out-of-plane magnetized magnetic tunnel junctions serve as storage elements provide both low switching currents, owing to the relatively weak Gilbert damping, and high thermal stability thanks to their large perpendicular magnetic anisotropy (PMA) values~[\cite{Ikeda10,Jan12,Dieny17}]. PMA was observed to be common at magnetic metal/oxide interfaces with either amorphous or crystalline oxides~[\cite{Rodmacq03,Manchon08,Nistor09}] and was theoretically attributed to the electronic hybridization between the oxygen and magnetic transition metal orbitals across the interface~[\cite{Yang11}]. 

The memory retention which is determined by the thermal stability of the storage layer magnetization is directly related to this layer's PMA. This is a key parameter in STT-MRAM applications. The concept of thermally assisted STT-MRAM was also proposed to reduce the write current while maintaining a large thermal stability factor~[\cite{Bandiera11}]. In regards to applications, the sensitivity of magnetic properties to temperature is a critical point especially for applications having to operate on a broad range of temperature such as the automotive ones. This emphasizes the need for a fundamental understanding of the PMA dependence on temperature. In this context, the correlation between temperature-dependent anisotropy constant $K$ and magnetization $M$ of ferromagnets, as described by Callen and Callen, follows a power scaling law: \mbox{$\frac{K(T)}{K(0)}=\left [\frac{M(T)}{M(0)}\right ]^{n}$} with $n=i(2i+1)$ corresponding to the $i$th order anisotropy constant~[\cite{Callen63,Callen66}]. It follows that for the first order anisotropy constant $K_{1}$, a scaling exponent $n=3$ is expected. However, deviations from this law with scaling exponents $n<3$ have been experimentally reported~[\cite{Gan11,Alzate14,Sato18}]. This was explained by the two-ion anisotropy model in L$1_0$ FePt alloys~[\cite{Mryasov05,Skomski06}] and more recently in CoFeB/MgO structures~[\cite{Sato18}]. However, considering various mechanisms that are likely to occur in actual devices and their probable contribution to the temperature dependence of the magnetic properties, the analysis of the relationship between thermal variations of anisotropy and magnetization requires further clarification. For instance, the effect of magnetic dead layer, which often forms when using buffer or capping layers for transition metal/oxide structures~[\cite{Cuchet14}], has not been addressed so far. Since the thickness of the dead layer is temperature dependent, considering its effect on the correlation between temperature dependences of $K$ and $M$ becomes essential. Another possible effect is the spatial thickness fluctuations at the interface~[\cite{Dieny94}] giving rise to a second-order anisotropy $K_{2}$ contribution manifested as a canted or easy cone magnetic states~[\cite{Timopheev16}]. All this emphasizes the importance of clarifying the mechanisms contributing to the temperature dependence of magnetic anisotropy and whether they are micro- or macroscopic in nature. In particular, the mechanisms underlying deviations from the theoretical Callen-Callen power law require further investigation.  

In this paper, we study theoretically the correlation between temperature dependences of magnetic anisotropy and magnetization in typical Fe/MgO-based structures addressing several fundamental and practical aspects. Our first-principles calculations carried out on ideal structure reveal that the second order magnetic anisotropy constant $K_{2}$ is one order of magnitude smaller than its first order one $K_{1}$ indicating that higher order anisotropy terms of intrinsic magnetocrystalline origin are unlikely to account for temperature dependent mechanisms. Neglecting higher-order anisotropy contributions to the atomistic spin Hamiltonian, calculations were performed in order to unveil the macroscopic mechanisms of temperature dependence of magnetic anisotropy. After calculating the layer-resolved exchange constants in ideal Fe/MgO interface, we demonstrate that including the effect of enhanced exchange at the interface is crucial to obtain scaling exponents consistent with Callen and Callen scaling power law. We elucidate that the layer-resolved temperature dependence of magnetic anisotropy also follows the law with $n=3$. Next, two mechanisms that are very likely to occur in magnetic metal/oxide structures are considered and their impact on the correlation between temperature dependences of $K$ and $M$ is analyzed. It is shown that the presence of a magnetic dead layer yields lower effective scaling exponents deviating from the theoretical law. The second mechanism reveals that increasing the percentage of inhomogeneity in the interfacial PMA decreases the scaling exponent. The present results elucidate that the mechanisms behind the deviations from the theoretical Callen-Callen power law are extrinsic (presence of a dead layer and of interfacial roughness) rather than intrinsic. These results allow us to coherently explain the difference in scaling exponents relating anisotropy and magnetization thermal variations reported in earlier experiments. 

The paper is organized as follows. In section II, we present the first-principles calculation results of the first and second order anisotropy constants at Fe/MgO interfaces with different terminations. The layer-resolved exchange constants are identified in section III. Then, the atomistic calculations of the temperature dependence of the magnetic anisotropy are presented and discussed in section IV. The effect of a magnetic dead layer is addressed in section V while section VI is dedicated to the effect of the inhomogeneity in the perpendicular magnetic anisotropy on the temperature dependence. Main conclusions drawn from this study are summed up in section VII.

\begin{figure}[ht]
 \centering
 \includegraphics[width=0.9\columnwidth]{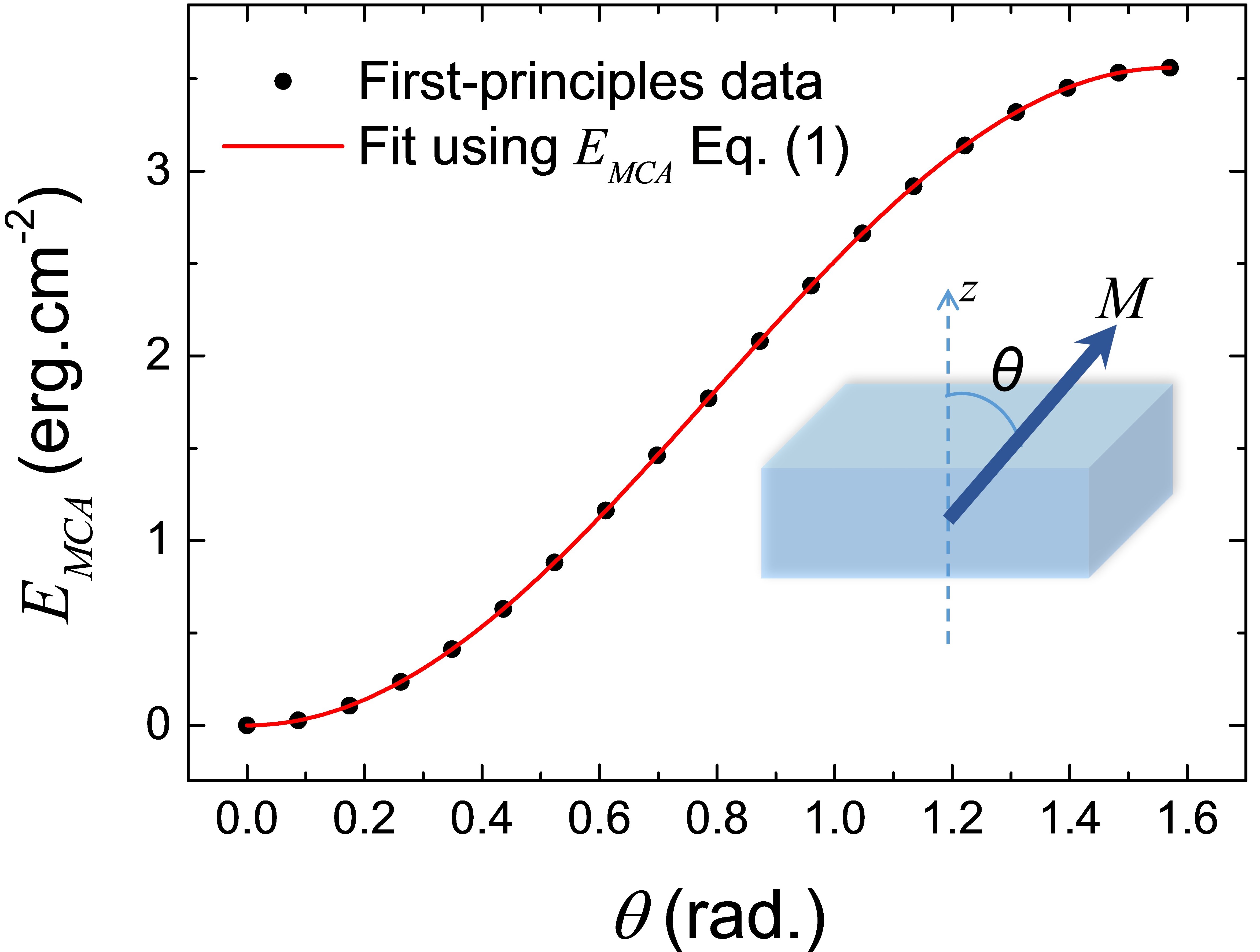}
\caption{(Color online) The magnetic anisotropy energy as a function of the angle $\theta$ between the magnetization direction and the normal to the interface calculated for Fe/MgO structure by first-principles (data points) is fitted to the $E_{MCA}$ Eq.~\ref{eq1} (solid line).}
\label{fig-fit-anisotropy}
\end{figure}

\begin{figure*}[ht]
 \centering
 \includegraphics[width=0.9\textwidth]{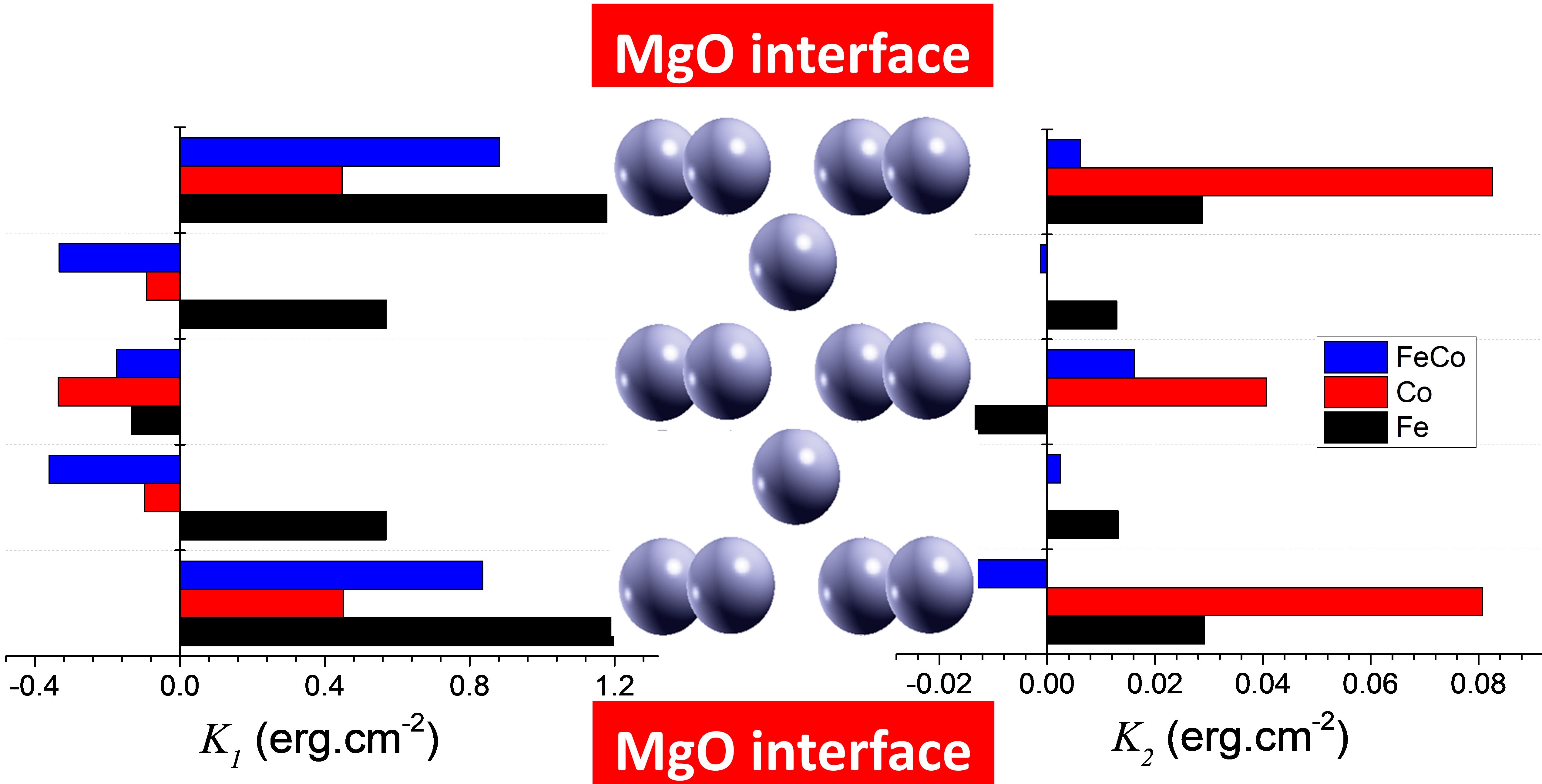}
\caption{(Color online) Layer-resolved values of $K_1$ and $K_2$, obtained from first-principles calculations, are compared for Fe/MgO and (Fe-Co)/MgO structures with either Co- or Fe-Co–terminated interfaces.}
\label{fig-K1-K2}
\end{figure*}

\section{$K_{1}$ and $K_{2}$ evaluation from first-principles}

The interfacial magnetic anisotropy energy $E_{MCA}$ is obtained from the dependence of the total energy $E$ on the angle $\theta$ between the magnetization direction and the normal to the interface expressed as:
\begin{equation}
\begin{split}
E_{MCA}(\theta)=E(\theta)-E(0) & = K_{1} sin^{2} \theta + K_{2} sin^{4} \theta \\
&+ K_{3} sin^6 \theta + ... = \sum K_{i} sin^{2i} \theta.
\end {split}
\label{eq1}
\end{equation}
Typically, the first order term $K_{1}$ dominates the higher order ones in structures with interfacial anisotropy. However, experiments have observed the influence of the second order $K_2$, namely in the formation of an easy-cone state~[\cite{Timopheev16,Teixeira18}]. The origin of significant $K_2$ values is attributed to spatial fluctuations of the film thickness~[\cite{Dieny94,Heinrich01}], strongly interface-concentrated PMA combined to a moderate exchange coupling of the interface moment to the rest of the film~[\cite{Sun15}] or strong magnetic inhomogeneities~[\cite{Timopheev17}]. Although aforementioned experiments were explained by such extrinsic mechanisms, it is necessary to elucidate whether the higher-order anisotropy terms may also be of intrinsic origin. For this, we performed systematic first-principles calculations to quantify the higher-order anisotropy terms in MgO-based interfaces including Fe/MgO and FeCo/MgO with either Co or FeCo termination. Details of the computation method are provided in the Appendix A.

\begin{table}[ht]
\centering
\begin{tabularx}{1\columnwidth}{l c c c c c c c c c}
 \hline
          &	 &	  & 	Fe/MgO	& 	 &	    &FeCo/MgO &	 &  	& FeCo/MgO  \\	
\hline
Termination&  &  & Fe        &    &     &Co             &   &      &FeCo           \\
\hline
$K_1$ (erg.cm$^{-2}$)&	 &	  &      $3.52$  &	 &  	&  $0.45$    &	 &	     &  $0.87$  \\
$K_2$ (erg.cm$^{-2}$)&	 &	  &   $0.038$	&	 &	     & $0.077$   & 	 &      & $0 $    \\
 \hline
\end{tabularx}
\caption{The first order $K_1$ and second order $K_2$ anisotropy constants obtained for Fe/MgO and (Fe-Co)/MgO structures with either Co- or Fe-Co–terminated interfaces by fitting the first-principles calculated $E_{MCA}$ to Eq.~ (\ref{eq1}).}
\label{tab1}
\end{table}

The total energies from first-principles calculations were fitted with Eq.~(\ref{eq1}) up to the third order. As an example, in Fig.~\ref{fig-fit-anisotropy} we show the result for the case of Fe/MgO structure. The obtained $K_1$ and $K_2$ values for the studied interfaces are summarized in Table~\ref{tab1}. One can observe that: (i) both $K_1$ and $K_2$ values are dependent on the interface termination while (ii) $K_2$ is found to be one order of magnitude smaller than $K_1$ and (iii) $K_{3}$ values are found to be negligible for all cases. 

\begin{figure}[h]
 \centering
 \includegraphics[width=\columnwidth]{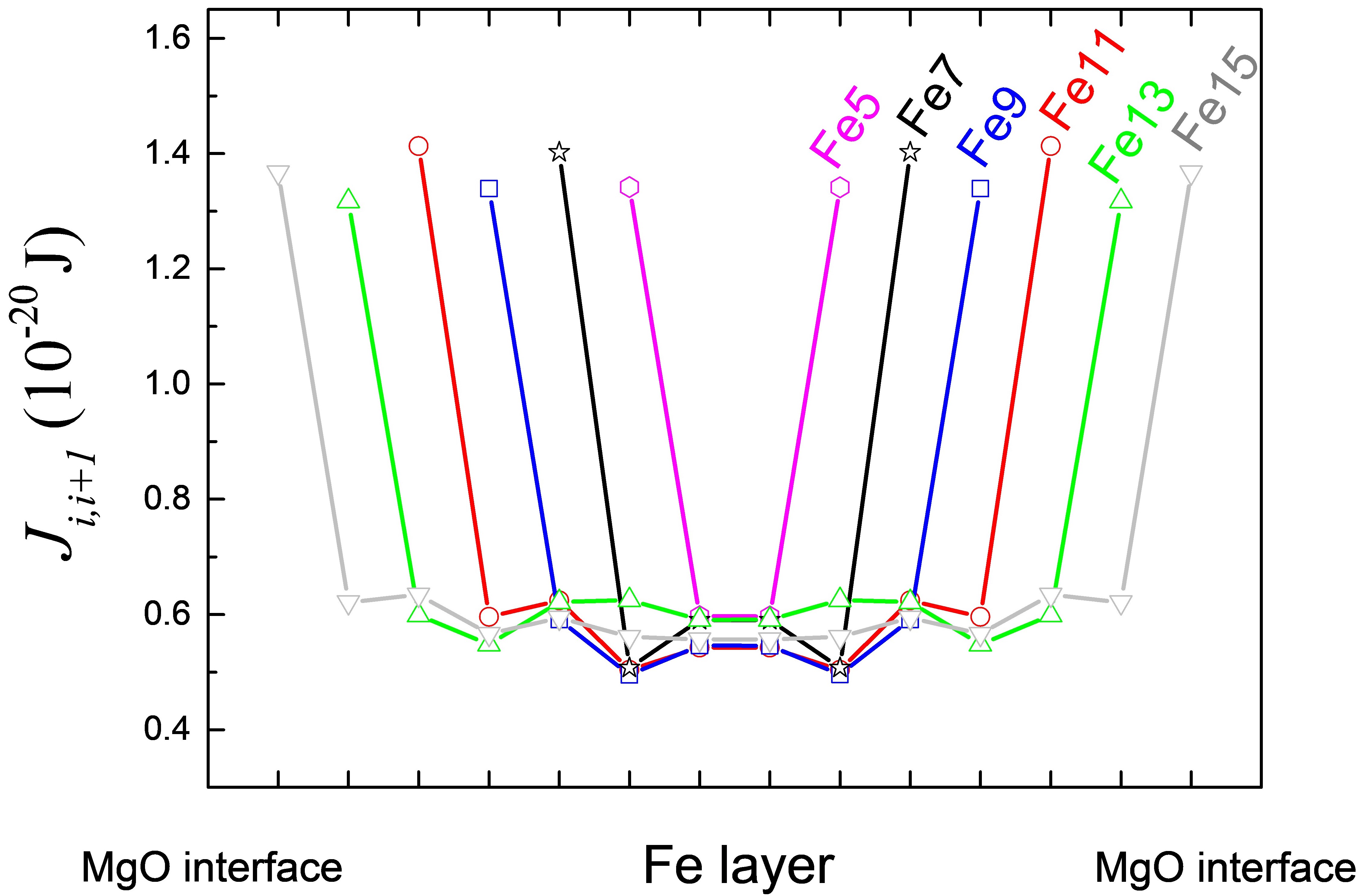}
\caption{(Color online) Layer-resolved nearest-neighbor exchange constants $J_{i,i+1}$ corresponding to the interaction between atom in layer $i$ and $i+1$ calculated from first-principles in MgO/Fe($t$)/MgO structures with Fe thickness varied between $5\le t \le 15$ monolayers (ML).}
\label{fig-exchange-parameters}
\end{figure}

\begin{figure*}[ht]
 \centering
 \includegraphics[width=0.85\textwidth]{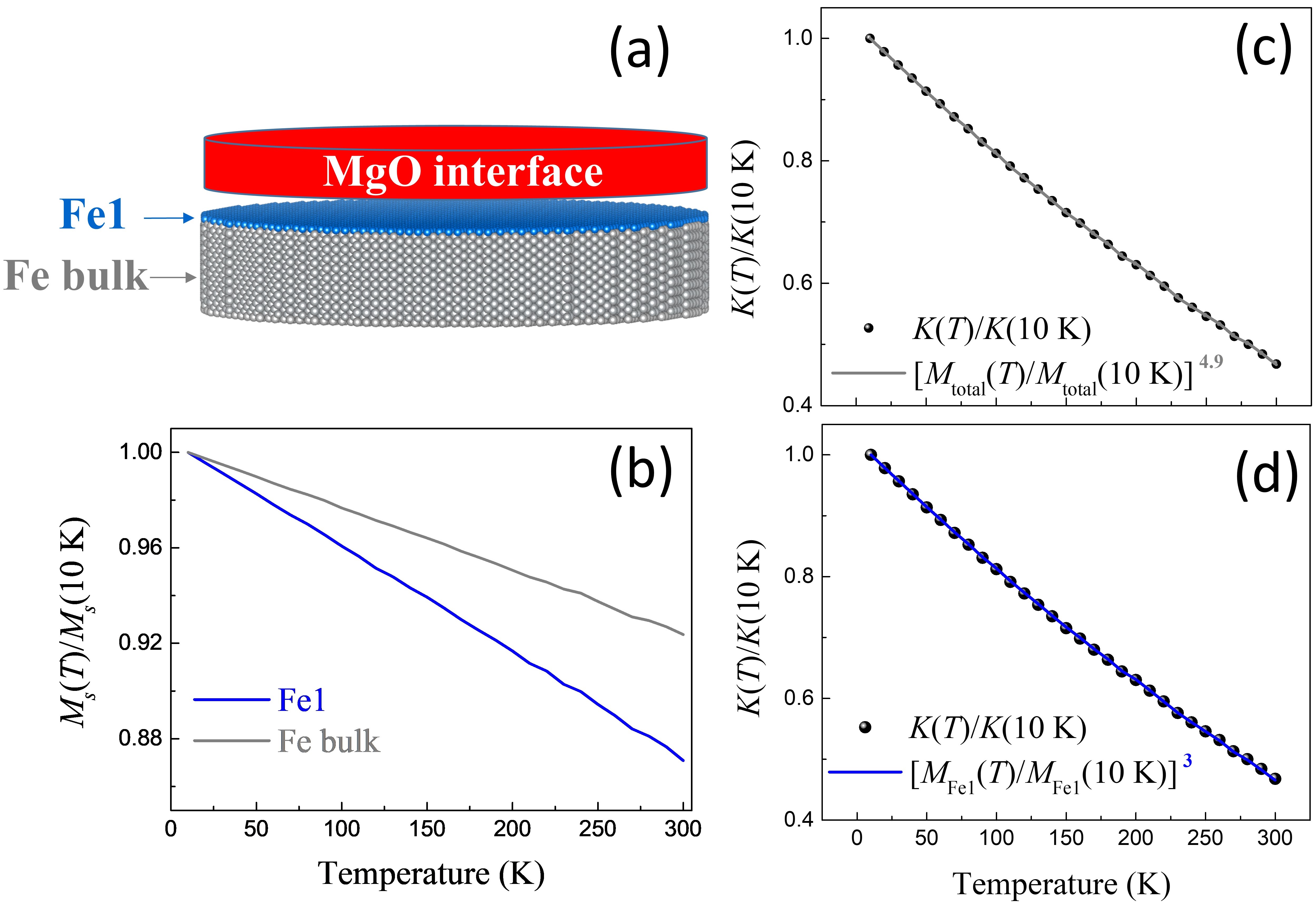}
\caption{(Color online) (a) The model Fe/MgO structure used to calculate the temperature-dependent magnetic properties comprising a bulk Fe region and one interfacial layer (Fe1). (b) The normalized saturation magnetization $M_s$ per region is plotted as a function of temperature. The normalized magnetic anisotropy $K$ as a function of temperature, shown with its fit to the normalized saturation magnetization of the whole structure, $M_{\rm{total}}$ (resp. the normalized saturation magnetization of the interfacial Fe1 layer $M_{Fe\rm{1}}$), yielding scaling exponent $n=4.9$ (resp. $n=3$), shown in panel (c) [resp. panel (d)].}
\label{fig-Fe1ML}
\end{figure*}

To get more insights, in Fig.~\ref{fig-K1-K2} we present the layer-resolved $K_1$ and $K_2$ values for different interfaces studied. In case of Fe/MgO $K_1$ is not entirely localized at the interface but rather propagates into the bulk showing an attenuating oscillatory behavior~\cite{Hallal13}. The layer-resolved $K_2$ for this structure follows the $K_1$ in trend and sign. The situation is different in the case of FeCo/MgO interface showing that for all terminations $K_1$ is overall lower compared to that at Fe/MgO and PMA originates from the first layer. As for the $K_2$, it is positive and one order of magnitude smaller than $K_1$ for Co-terminated interface, while it is negligible and depends strongly on the atomic distribution for FeCo-terminated interface. It is noteworthy to mention that the asymmetry in the structure between the two interfaces in the case of FeCo-terminated is at the origin of the observed asymmetry in the $K_{2}$ values (c.f. Appendix A (1)).   

Based on these findings, we conclude that significant $K_2$ are not of intrinsic origin in MgO-based interfaces and its emergence should be attributed to extrinsic mechanisms. In particular, fluctuations of interfacial PMA were pointed out by experiments~\cite{Timopheev17} and modeled analytically~\cite{Heinrich01} or by micromagnetic simulations~\cite{Mohammadi18} giving rise to significant $K_2$ values. Thus, in order to address the temperature dependence of magnetic anisotropy, we exclude $K_2$ from the atomistic spin Hamiltonian. In the following, we focus instead on exploring two extrinsic mechanisms that are commonly encountered in experiments, namely the presence of a magnetic dead layer and of spatial inhomogenities in the interfacial PMA.

\section{Calculation of the exchange constants}

To study the temperature dependence of magnetic anisotropy, it is important to determine the fundamental parameters that are decisive for this effect. As presented in the Calculation Methods, the first-principles calculations provide both the layer-resolved magnetic moments and anisotropy values at zero temperature. We notably observe the enhancement of those two parameters at the two Fe interfacial layers compared to their bulk counterparts in agreement with previous reports~\cite{Yang11,Hallal13}. As for the exchange constants, one might expect reduced values at surfaces or interfaces due to the reduced atomic coordination based on a classical Heisenberg model. However, an enhancement of the exchange interaction was reported at different magnetic surfaces using ab initio calculations~\cite{Turek2003}. Thus, obtaining layer-resolved exchange constants is important not only for characterizing the Fe/MgO interface but also to study macroscopic effects such as magnetic dead layers created by atomic inter-diffusion at the Fe/Ta interface as discussed later.

Using first-principles calculations, we calculated the long-ranged exchange constants in MgO/Fe($t$)/MgO structures with variable Fe thickness, $5 \le t$ (ML)$ \le 15$. Since the exchange interaction is strongly distance-dependent, truncating it to include only nearest-neighbors interactions remains a good approximation when used in parameterizing atomistic simulations so that to reduce the computational effort. As explained in the Calculation Methods, this is done by including the tail of the long-ranged exchange into the nearest-neighbor one so that the exchange interaction between atomic monolayers remains same. The layer-resolved nearest-neighbor exchange pairs $J_{i,i+1}$ between an atom located at layer $i$ and another at $i+1$ are shown in Fig.~\ref{fig-exchange-parameters} for variable Fe thickness. It can be clearly seen that the values of the exchange interaction at the interface ($J_{i,i+1}\approx 1.3-1.4 \times 10^{-20}$ Joule) are well enhanced compared to their bulk counterpart values ($J_{i,i+1}\approx 5-6  \times 10^{-21}$ Joule) independently of the Fe thickness. This finding is consistent with enhanced exchange interaction reported at Fe(001) surface~\cite{Turek2003}. We also note that the presence of quantum well states leads to the observed oscillations in the bulk exchange values as a function of the distance from the interface~\cite{Milun2002}.

\section{Total and layer-resolved temperature dependence of magnetic anisotropy in ideal Fe/MgO interface}

Before investigating the macroscopic mechanisms that contribute to the temperature dependence of magnetic anisotropy, let us first address its total and layer-resolved behavior for ideal Fe/MgO interface. The details of the calculations are provided in the Appendix A(2). We start with a simple model using uniform exchange fixed to the Fe bulk exchange constant ($J=7\times10^{-21}$ J~\cite{Evans14}). The 15 ML thick system comprises one Fe layer with interfacial magnetic anisotropy (Fe1) and the bulk region with negligible anisotropy [Fig.~\ref{fig-Fe1ML}(a)]. The normalized saturation magnetization $M_s$ of Fe1 decreases more rapidly with temperature compared to that of the bulk as shown in Fig.~\ref{fig-Fe1ML}(b) due to the reduced atomic coordination at the interface. Next, we fit and compare the calculated anisotropy $K$(T) to $M_s$(T) in two cases. If the total saturation magnetization $M_{total}$(T) is considered, we obtain a scaling exponent $n=4.9$ [Fig.~\ref{fig-Fe1ML}(c)]. However, using the interface saturation magnetization $M_{Fe1}$(T) yields $n=3$, i.e. \mbox{$\frac{K_{Fe1}(T)}{K_{Fe1}(10 K)}=\left [\frac{M_{Fe1}(T)}{M_{Fe1}(10 K)}\right ]^3$} following exactly the Callen-Callen law [Fig.~\ref{fig-Fe1ML}(d)].

\begin{figure}[h]
	\centering
	\includegraphics[width=\columnwidth]{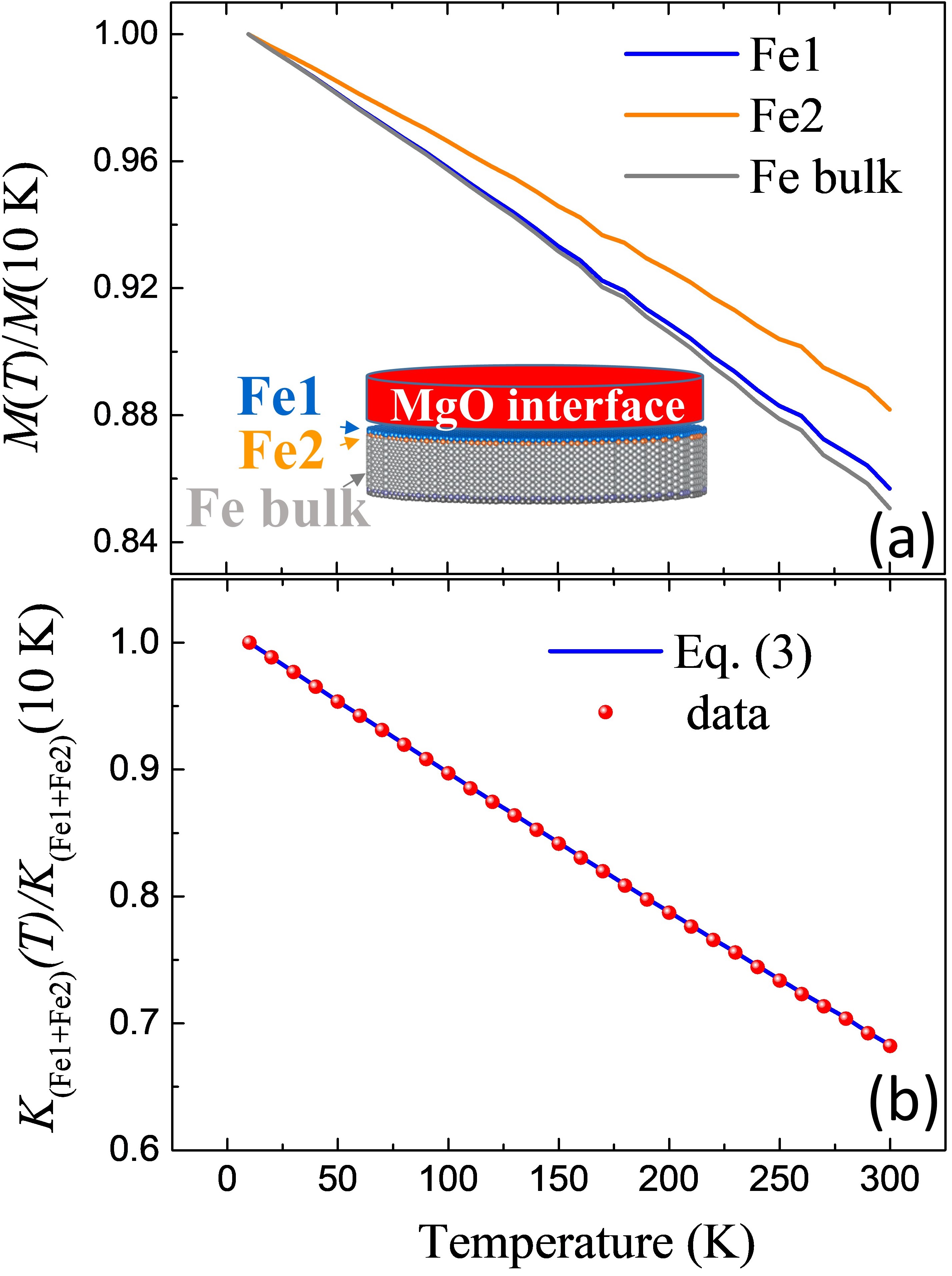}
	\caption{(Color online) (a) The normalized saturation magnetization $M_s$ per region as a function of the temperature for the model Fe/MgO structure used to calculate the temperature dependent magnetic properties comprising a bulk Fe region and two interfacial layers (Fe1 and Fe2). (b) The calculated normalized magnetic anisotropy $K_{Fe1+Fe2}$ as a function of temperature is shown by circle data points. The solid line corresponds to Eq.~\ref{eq3}.}
	\label{fig-Fe2ML}
\end{figure}

Next, we improve the model by including the layer-resolved exchange constants calculated from first-principles while keeping the magnetic anisotropy associated to Fe1 only. In this case, by fitting $K$(T) to $M_{total}$(T) a scaling exponent $n=3$ is obtained. This emphasizes the importance of including the enhanced interfacial exchange in the model to obtain reasonable scaling exponents. We further improve our model, by including the layer-resolved magnetic anisotropy of the two interfacial Fe layers Fe1 and Fe2. Fig.~\ref{fig-Fe2ML}(a) compares the layer-resolved saturation magnetization for the three different regions. The magnetization of Fe2 reduces with temperature slower than that of Fe1 and bulk region. This can be explained by the enhanced exchange at the interface, namely in $J_{1,2}$ nearest-neighbor pair interaction, compared to bulk values accompanied by a reduced atomic coordination of the Fe1 layer and a full coordination of Fe2. This has a direct implication on the scaling exponent which is found to be $n=2.8$ by fitting $K$(T) to $M_{total}$(T).

To extract per layer resolved correlation between $K(T)$ and $M_{s}(T)$, the results of previous two models are used. Considering two interfacial Fe monolayers, the expression of the temperature-dependent anisotropy can be expanded as: 
\begin{widetext}
\begin{equation}
\frac{K_{Fe1+Fe2}(T)}{K_{Fe1+Fe2}(10K)}=\frac{K_{Fe1}(T)}{K_{Fe1}(10K)}\left [\frac{K_{Fe1}(10K)}{K_{Fe1}(10K)+K_{Fe2}(10K)}\right ]+\frac{K_{Fe2}(T)}{K_{Fe2}(10K)}\left [\frac{K_{Fe2}(10K)}{K_{Fe1}(10K)+K_{Fe2}(10K)}\right ].
\label{eq2}
\end{equation}
\end{widetext}
Assuming that the magnetic anisotropy of every Fe layer $i$ ($K_{Fe_i}$) scales as $n=3$ with its saturation magnetization $M_{Fe_i}$ according to the Callen-Callen power law, Eq.~(\ref{eq2}) can be rewritten as
\begin{widetext}
\begin{equation}
\frac{K_{Fe1+Fe2}(T)}{K_{Fe1+Fe2}(10K)}=\left [\frac{M_{Fe1}(T)}{M_{Fe1}(10K)}\right ]^3 \left [\frac{K_{Fe1}(10K)}{K_{Fe1}(10K)+K_{Fe2}(10K)}\right ]+\left [\frac{M_{Fe2}(T)}{M_{Fe2}(10K)}\right ]^3\left [\frac{K_{Fe2}(10K)}{K_{Fe1}(10K)+K_{Fe2}(10K)}\right ].
\label{eq3}
\end{equation}
\end{widetext}
The calculated data accounting for the anisotropy of both Fe1 and Fe2 that yields $n=2.8$ are then compared to those obtained using Eq.~\ref{eq3} [Fig.~\ref{fig-Fe2ML}(b)]. The excellent agreement between the two curves confirms the validity of our assumption that the layer-resolved behavior follows the Callen-Callen scaling power law for each individual Fe layer: $\frac{K_{Fe_i}(T)}{K_{Fe_i}(10 K)}=\left [\frac{M_{Fe_i}(T)}{M_{Fe_i}(10 K)}\right ]^3$.    

\begin{figure}[ht]
	\centering
	\includegraphics[width=\columnwidth]{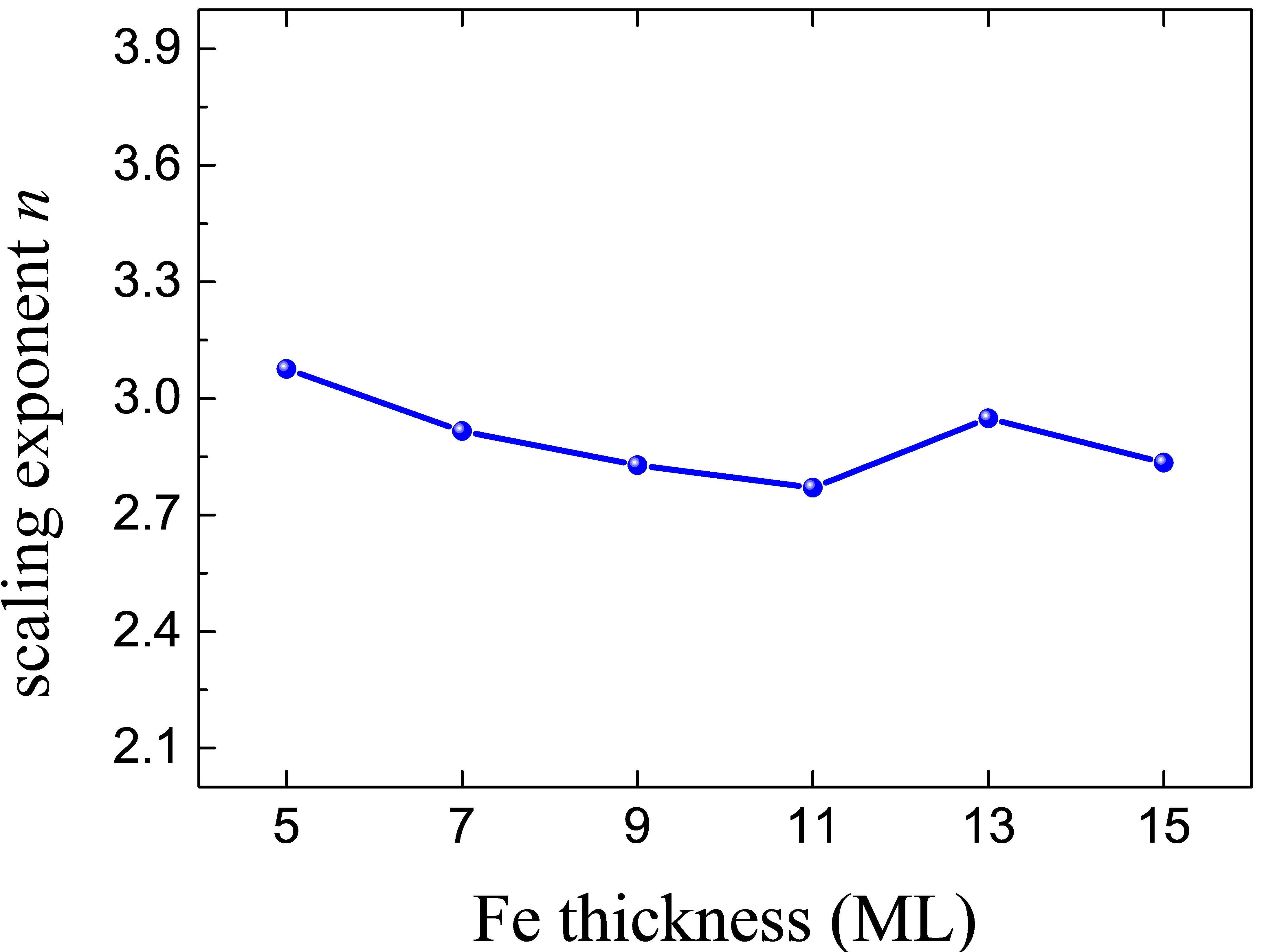}
	\caption{(Color online) The scaling exponent $n$ of the temperature dependence of magnetic anisotropy for Fe/MgO interface calculated as a function of the Fe thickness.}
	\label{fig-exponent-vs-thickness}
\end{figure}

To examine the effect of the sample thickness on the scaling exponent, we have used our previously calculated layer-resolved exchange constants for Fe thickness varying as $5\le t \le 15$ MLs. The scaling exponents obtained by fitting $K$(T) to $M_ {total}$(T) as a function of Fe thickness are shown in Fig.~\ref{fig-exponent-vs-thickness} and range between $2.8 \le n \le 3.1$. This demonstrates that Callen-Callen power law holds for the anisotropy of Fe/MgO interfaces independently of the sample thickness. This is remarkable, given that the enhanced exchange at the interface is crucial to get reasonable values of the scaling exponent around $3$. Based on all this, one can conclude that an ideal Fe/MgO interface with its intrinsic properties can not explain the observed deviations from the Callen-Callen power law. This finding indicates that the deviations may be due to macroscopic effects that we will address in the remainder of this paper.

\section{Effect of magnetic dead layer}


\begin{figure}[t]
	\centering
	\includegraphics[width=\columnwidth]{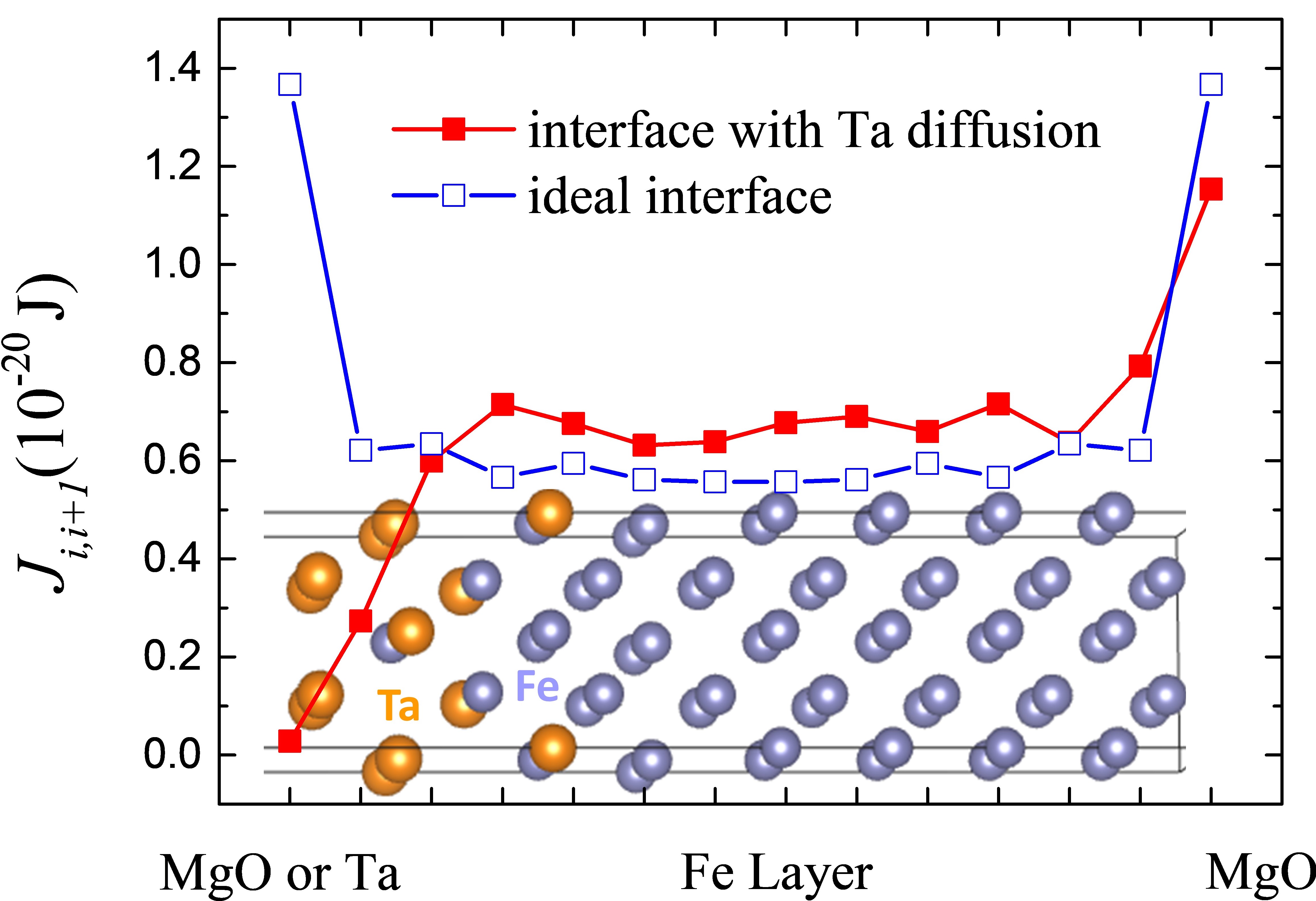}
	\caption{(Color online) The calculated nearest-neighbor exchange constants across this system are represented by closed red squares and compared to the values in an ideal Fe/MgO interface represented by open blue squares. The inset displays the structure used to model Ta diffusion in a 15 ML thick Fe/MgO. Ta atoms are introduced in the four outermost layers opposite to the MgO with 100, 75, 50, and 25\% respectively.}
	\label{fig-dead-layer-exchange}
\end{figure}

\begin{figure}[t]
	\centering
	\includegraphics[width=\columnwidth]{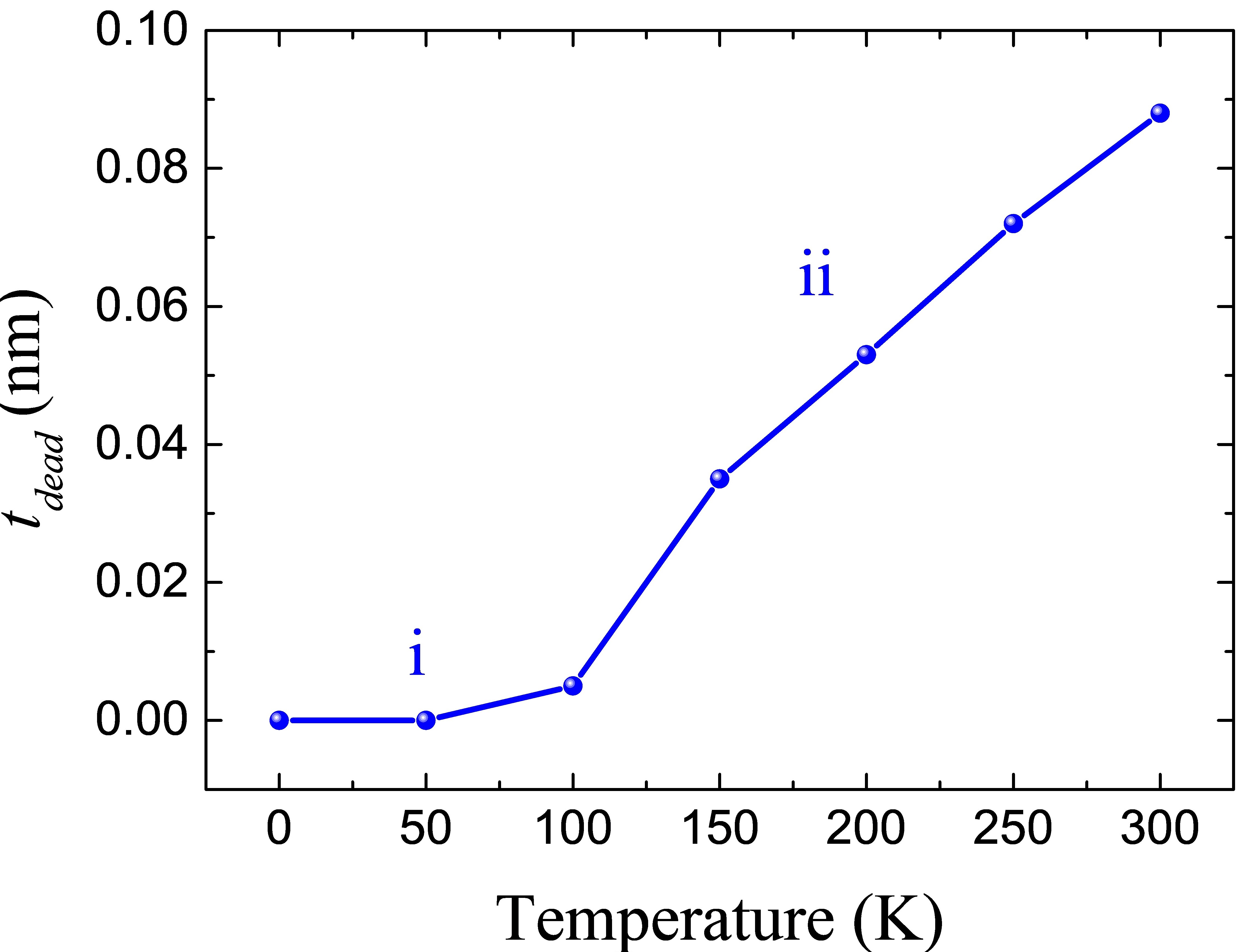}
	\caption{(Color online)  The variation of magnetic dead layer thickness $t_{dead}$ plotted as a function of temperature shows two regimes: a negligible thickness regime denoted by (i) and another revealing an increase of $t_{dead}$ with temperature (ii).}
	\label{fig-tdead-vs-T}
\end{figure}

Magnetic dead layers form when the atoms of the capping or buffer layer diffuse into the magnetic region upon annealing resulting in a reduced coordination. We therefore chose to model the dead layer by introducing Ta diffusion in the outermost four Fe layers opposite to the Fe/MgO interface with respective Ta percentage of 100, 75, 50, 25\% [see the inset of Fig.~\ref{fig-dead-layer-exchange}]. \footnote{Ta is chosen since it is widely used in experiments. We point out that the effect of the buffer layer is beyond the scope of the present study which focuses on the mechanisms of the temperature dependence, however it is an interesting issue that is worth to be addressed in the future.} After relaxing the structure, the nearest-neighbor exchange constants were calculated and are compared to those of an ideal Fe/MgO interface in Fig.~\ref{fig-dead-layer-exchange}. Using those exchange constants, atomistic calculations for variable system thickness were performed in order to extract the dead layer thickness. We find that the estimated dead layer thickness $t_{dead}$ is negligible at sufficiently low temperatures [region (i) in Fig.~\ref{fig-tdead-vs-T}] since the Fe layers hosting Ta atoms still contribute a substantial saturation magnetization $M_s$ (see Appendix C). However, increasing the temperature results in the decrease of their $M_s$, which in turn increases the dead layer thickness [region (ii) in Fig.~\ref{fig-tdead-vs-T}].

\begin{figure}[t]
	\centering
	\includegraphics[width=\columnwidth]{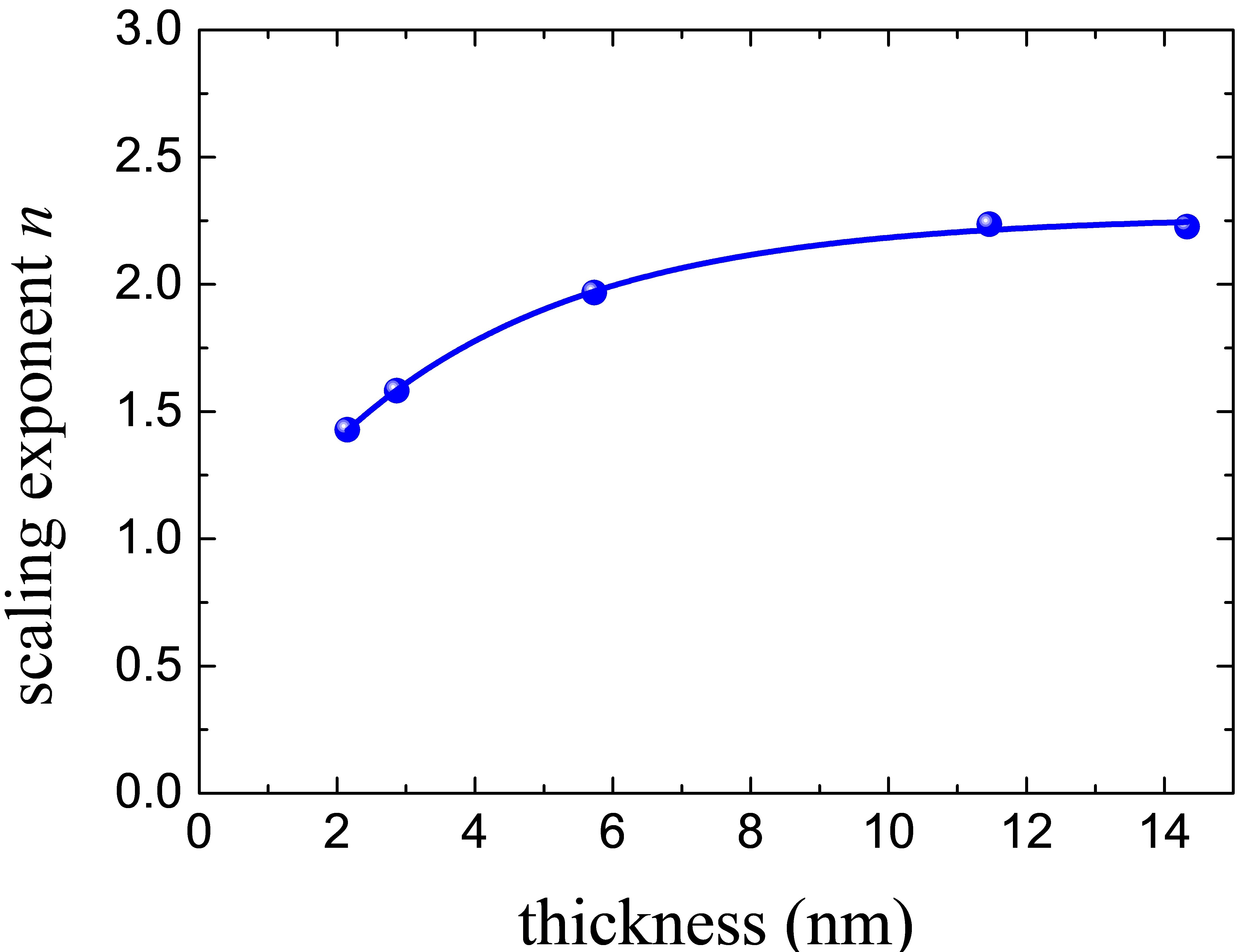}
	\caption{(Color online) The effect of the magnetic dead layer on the temperature dependence of $K$ is revealed by the variation of the scaling exponent $n$ as a function of the magnetic layer thickness.}
	\label{fig-exponent-dead-layer}
\end{figure}

In order to check how the magnetic dead layer affects the correlation between temperature dependence of $K$ and $M_s$, we calculated the scaling exponent as a function of thickness of the system presented in Fig~\ref{fig-exponent-dead-layer} whose variation can be fitted exponentially. Interestingly, the scaling exponent for all considered thicknesses significantly deviates from the Callen-Callen power law and ranges between $1.5 \le n \le 2.2$. Thus, the low values of the experimental scaling exponents reported earlier can be attributed to the presence of a magnetic dead layer. Noteworthy, modeling magnetic dead layers in this way provides a qualitative insight of its influence on the temperature dependence of the magnetic properties of realistic samples. In other words, changing the model of the Ta diffusion would modify the exchange constants and consequently the values of the scaling exponents without changing the trend of its variation with the sample's thickness.

\section{Effect of inhomogeneity in perpendicular magnetic anisotropy}

\begin{figure}[b]
	\centering
	\includegraphics[width=\columnwidth]{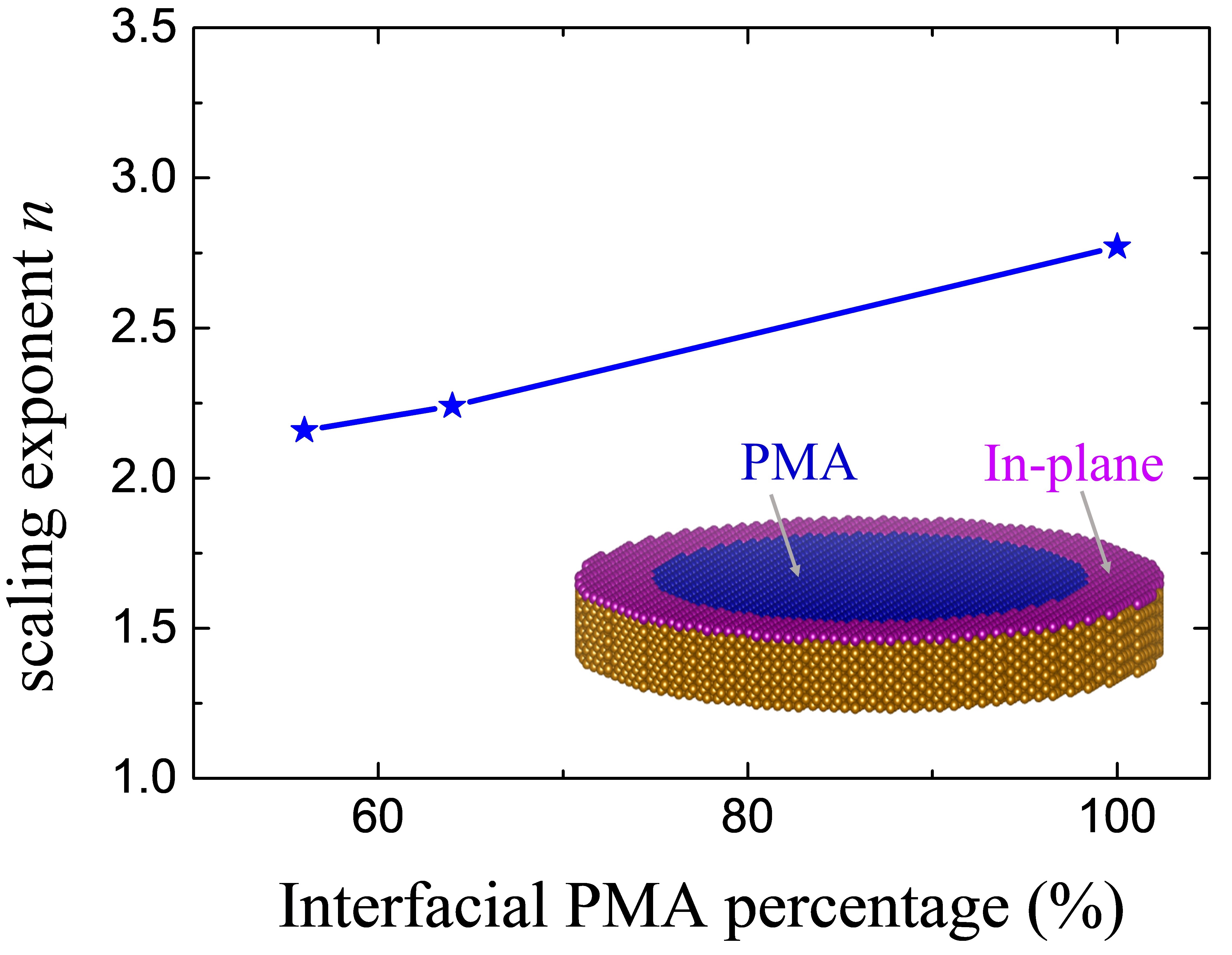}
	\caption{(Color online) The variation of the scaling exponent $n$ is plotted as a function of the percentage of intefacial PMA obtained by employing the core shell model shown in the inset.}
	\label{fig-core-shell}
\end{figure}

As we have already mentioned above, an analytical model~\cite{Dieny94,Heinrich01} and recent experiments~\cite{Timopheev16,Timopheev17} pointed out the presence of spatial fluctuations in the interfacial magnetic anisotropy. These fluctuations were attributed to local variations in the ferromagnetic layer thickness associated with film roughness or to the diffusion of buffer layer atoms through the ferromagnetic layer towards the MgO interface upon post-deposition annealing. Consequently, it is important to investigate the effect of inhomogeneous interfacial perpendicular magnetic anisotropy on its temperature dependence. We thus employ the core shell model comprising two regions at the interface, one with a perpendicular (PMA) and the other possessing in-plane magnetic anisotropy, whose values are equal in magnitude but opposite in sign, as shown in the inset of Fig.~\ref{fig-core-shell}. The resulting dependence of the scaling exponent on the percentage of the interfacial PMA is shown in Fig.~\ref{fig-core-shell}. One can see that the scaling exponent decreases to values below $2.5$ as the interfacial PMA percentage decreases. This highlights that the effect of interfacial roughness plays also a crucial role in the temperature dependence of the magnetic properties of MgO-based interfaces and leads to deviations from the theoretical Callen-Callen limit.

\section{Conclusion}

The results presented in this work unveiled several important mechanisms for understanding temperature dependence of magnetic anisotropy at MgO-based interfaces. Based on small second order anisotropy $K_2$ values obtained from the first-principles calculations on ideally homogeneous interfaces, we concluded that higher-order anisotropy terms observed in some experiments arise mainly from extrinsic contributions such as spatial fluctuations of first order anisotropy. Therefore, only the first order $K_1$ term was included in the atomistic spin Hamiltonian to describe the temperature dependence of the magnetic anisotropy. First principles calculations of the layer-resolved exchange constants have shown an enhancement of the values at the Fe/MgO interface compared to the bulk which provided a more accurate description of the temperature dependence of magnetic properties in our atomistic model.  In an ideal Fe/MgO interface, the temperature-dependence of the total and layer-resolved anisotropy follow the Callen and Callen scaling power law and thus intrinsic properties can't explain deviations from this law. In this respect, we have shown that such deviations observed experimentally can be attributed to two macroscopic mechanisms: the presence of a magnetic dead layer and the spatial fluctuations of the interfacial PMA. In particular,   the scaling exponent was found to vary with the sample's thickness in the presence of a magnetic dead layer. As a matter of fact, this dependence was observed in Ref.~\cite{Bandiera-thesis} at Co/AlOx interface.  Distinguishing between these two mechanisms experimentally might be possible by either of the following two ways: (i) Thickness-dependent measurements can be performed knowing that the mechanisms yield different thickness-dependence of $n$; (ii) Interface-sensitive techniques such as M{\"o}ssbauer experiments with Fe doped interfacial layer can give access to the interfacial magnetization. Accordingly, if scaling $K(T)$ by the interfacial magnetization only leads to $n=3$ independently of the magnetic layer thickness, then a magnetic dead layer mechanism could be concluded. Otherwise,still considering the interfacial magnetization only, $n<3$ scaling exponents indicate an inhomogeneous interfacial PMA mechanism. We anticipate that the provided description of temperature dependence of magnetic properties at MgO-based interfaces will help for fundamental and technical understanding of thermal effects.  
    
We thank R. F. L. Evans for helpful discussions and acknowledge partial support by the Advanced Storage Research Consortium (ASRC) and by the European Research Council (ERC) Advanced Grant Project MAGICAL No. 669204.


\appendix
\renewcommand{\thesection}{\Alph{section}}
\setcounter{section}{0}


\section{Calculation Methods}

\subsection{First-principles calculations}
\begin{figure}[ht]
	\centering
	\includegraphics[width=0.95\columnwidth]{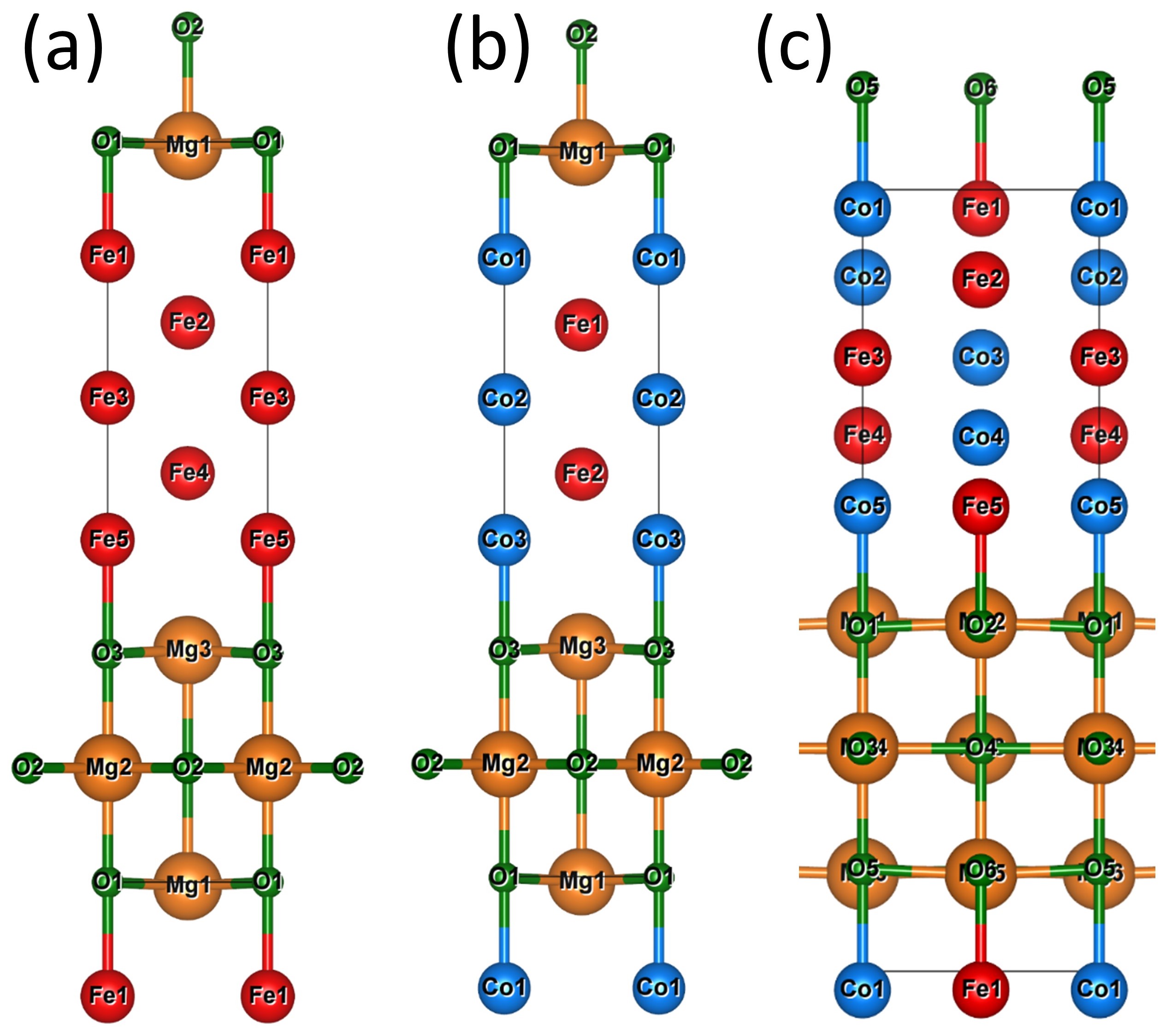}
	\caption{(Color online) The periodic structures used in the first-principles calculations. (a) Fe-, (b) Co-, and (c) FeCo-terminated MgO-based interfaces.}
	\label{fig-structures}
\end{figure}

Our first-principles calculations are based on the projector-augmented wave (PAW) method \cite{Blochl} as implemented in the VASP package \cite{Kresse93,Kresse96,Kresse3} using the generalized gradient approximation \cite{Perdew96} and including spin-orbit coupling within the second-order perturbation theory~\cite{Laan98}. A kinetic energy cutoff of $550$ eV has been used for the plane-wave basis set and a $31\times 31\times 1$ $k-$point mesh to sample the first Brillouin zone. The periodic structures comprises five magnetic monolayers (ML), Fe or FeCo, sandwiched between five ML of MgO as shown in the Fig.~\ref{fig-structures}. As known for Fe/MgO interface, the most stable location for the oxygen atoms is on top of metal ions due to the strong overlap between Fe-3d and O-2p orbitals~\cite{Yang11}. For CoFe/MgO interface, CoFe has a CsCl structure and it is in B2-ordered phase with the CoFe (100) parallel to the MgO (110) direction where we have considered two cases. The first is with Co-termination, as shown in Fig.~\ref{fig-structures}(b), which was found to be the more stable in previous reports~\cite{Burton2006,Khoo13}. The second case is with 50\% Fe-50\% Co termination, as shown in Fig.11 (c), where the transition metal atoms are located above the O atoms at the interface. All the structures were fully relaxed in atomic positions and volume until the forces became smaller than $1$ meV/\AA{}. We found an optimized lattice constant $a= 2.976$, $2.918$, and $4.149$ \AA{} for the cases of Fe/MgO, Co-terminated FeCo/MgO, and FeCo-terminated FeCo/MgO, respectively. The layer-resolved magnetic anisotropy contributions are evaluated following \cite{Hallal13}. 

The calculations of the exchange constants are done in two steps. First, first-principles calculations based on linear combination of atomic orbitals (LCAO) were performed using the SIESTA package~\cite{Soler2002} within the PBE form of the exchange-correlation functional~\cite{Perdew96}. A linear combination of numerical atomic orbitals with a double-$\zeta$ polarized basis set, a real-space grid cutoff of $1000$ Ry, and $25\times 25 \times 3$ $k$-point sampling of the Brillouin zone were used. In the second step, the converged Hamiltonian and overlap matrix are used by the TB2J package~\cite{He2021} to calculate the long-ranged exchange interactions. This is done based on the Green’s function method and using the magnetic force theorem which takes the rigid spin rotation as a perturbation to the electronic structure. To perform atomistic calculations, the short-ranged exchange constants are obtained from the long-ranged ones by truncating to the nearest-neighbor interactions such that the exchange between an atom located at layer $i$ and another at $i+1$ is calculated by summing over all atomic planes above and below layer $i$ where $J_{i,i+1}=\frac{1}{4}\sum_{n\le i,m>i}J_{n,m}$. Noteworthy, we have performed additional TB-LMTO calculations~\cite{Questaal} of the exchange interaction that showed similar trends with Fe thickness.

\begin{figure}[h]
	\centering
	\includegraphics[width=0.95\columnwidth]{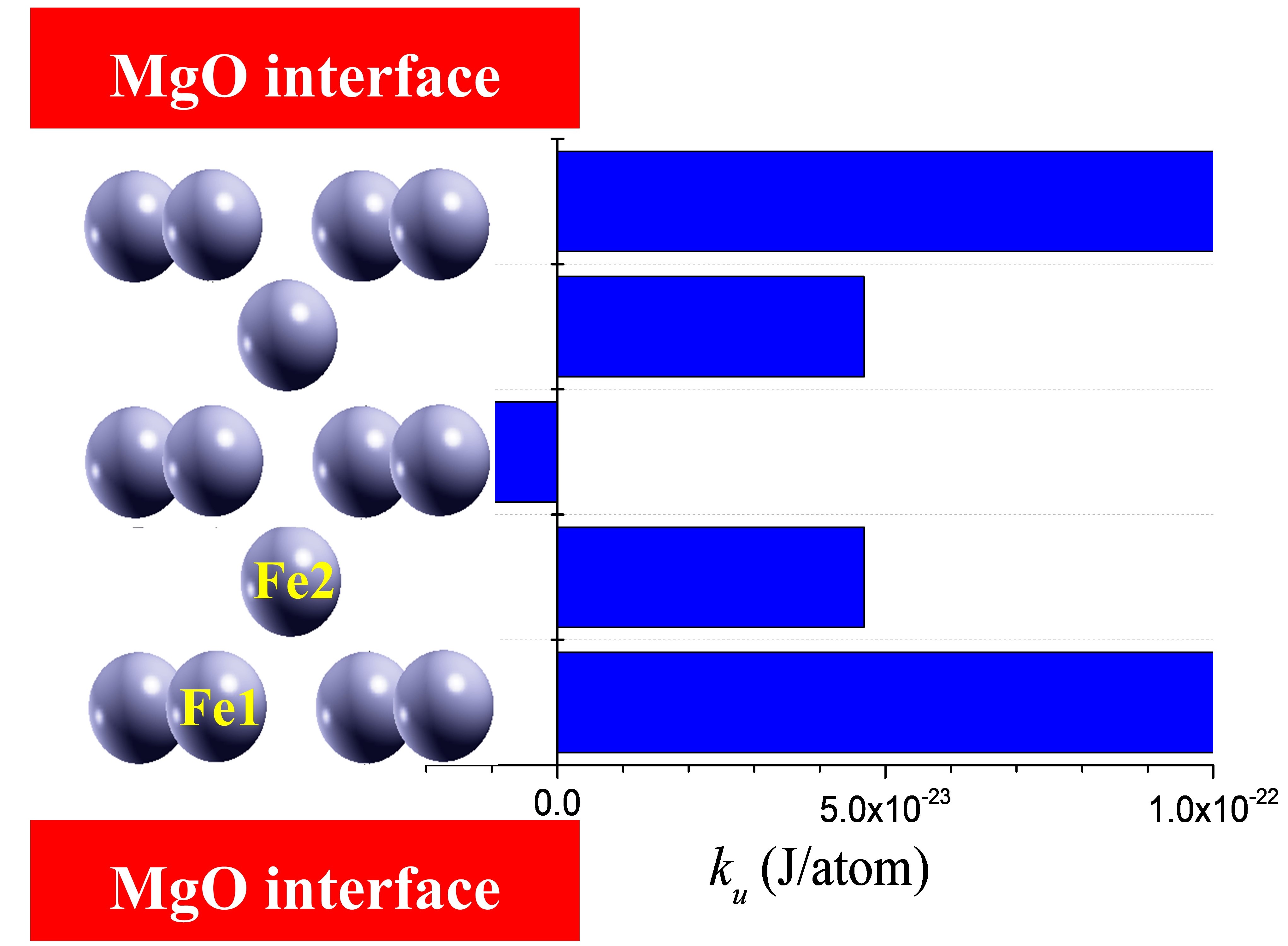}
	\caption{(Color online) Layer-resolved anisotropy energies to first order $E_{MCA}$ at Fe/MgO interface, obtained from first-principles calculations, showing that the both the first (Fe1) and second (Fe2) Fe monolayers contribute to the PMA. Those values are used to parameterize the temperature-dependent atomistic calculations.}
	\label{fig-layered}
\end{figure}

\begin{figure}[hb]
	\centering
	\includegraphics[width=0.95\columnwidth]{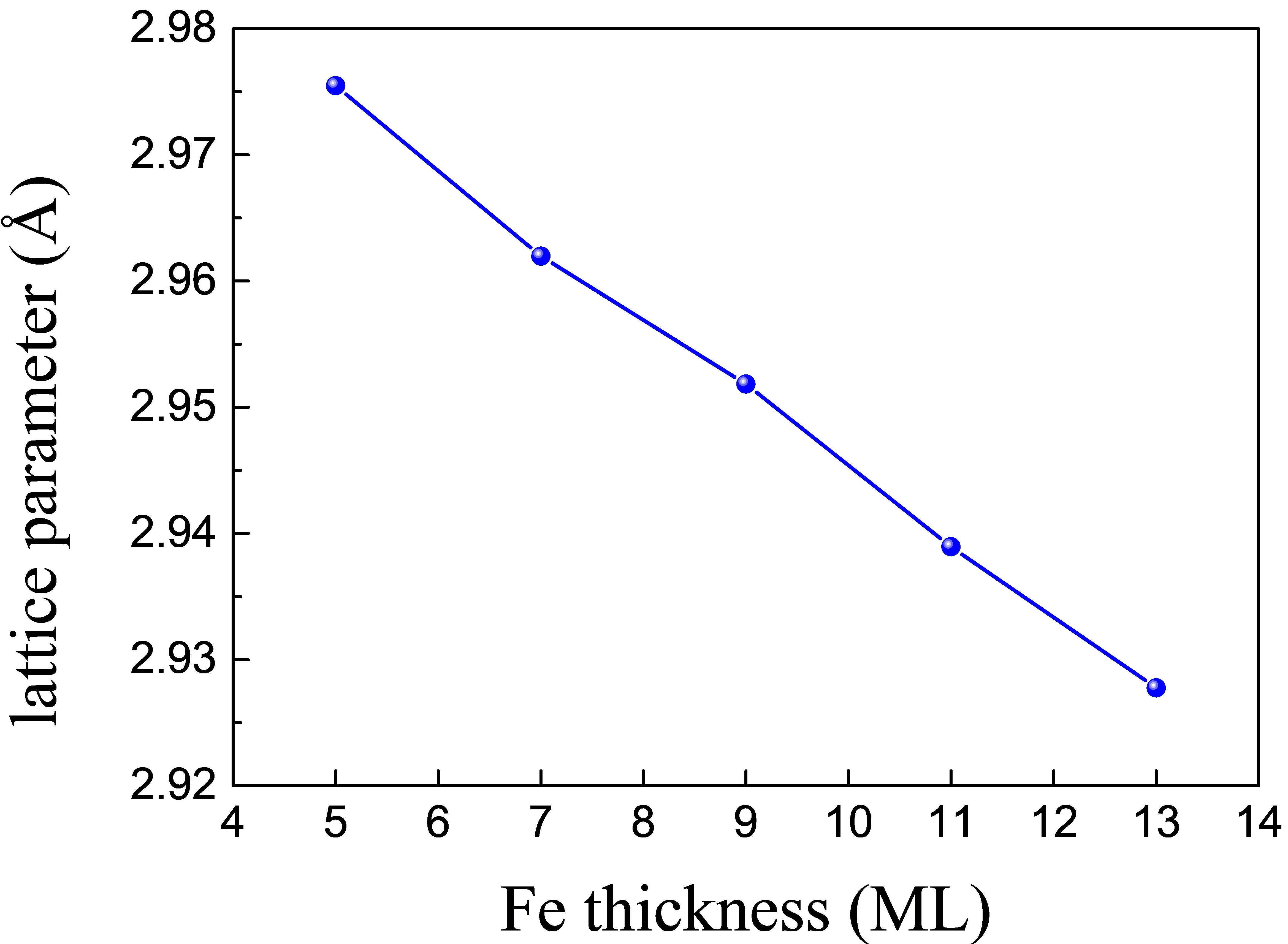}
	\caption{(Color online) The optimized in-plane lattice parameter of Fe/MgO supercell varies with Fe thickness.}
	\label{fig-app3}
\end{figure}

\subsection{Atomistic calculations}
The atomistic calculations are based on the classical atomistic spin model implemented in the VAMPIRE package~\cite{Evans14,Vampire}. This allows to study the influence of thermal spin fluctuations on the intrinsic magnetic properties such as the magnetization and magnetic anisotropy. The spin Hamiltonian including the exchange interaction and the uniaxial magnetic anisotropy is written as:
\begin{equation}
 H=-\sum_{i<j}J_{ij}{\bf S}_{i} . {\bf S}_{j} - \sum_{i} k_{u}({\bf S}_{i} . {\bf e}_{i})^{2} ,
\label{eqA1}
\end{equation}
where $J_{ij}$ is the isotropic exchange constant between nearest-neighboring atomic sites of local spin moment directions ${\bf S}_{i}$ and ${\bf S}_{j}$, $k_{u}$ is the uniaxial anisotropy constant per atom, and ${\bf e}_i$ is a unit vector along the magnetic easy axis. Only the first order anisotropy constant is included in the Hamiltonian since our first-principles calculations demonstrated small second order $K_2$ values. To model the Fe/MgO interface, we construct a cylindrical-shaped system with a $15$ nm diameter and we vary the thickness $t$ from the body-centered cubic (bcc) Fe crystal of lattice parameter $a=2.866$ \AA{}. We fixed the in-plane lattice parameter in all calculations since we found that it has a negligible effect on the calculated properties (Appendix B). We divide the system into two regions: (i) bulk with negligible $k_{u}$ and atomic spin moment of $2.2$ $\mu_{B}$, and (ii) interfacial region with an enhanced atomic spin moment of $2.76$ ($2.49$) $\mu_{B}$ and $k_{u}=1.099 \times 10^{-22}$ ($4.593 \times 10^{-23}$) J/atom for the first Fe1 (second Fe2) monolayer, respectively. Those site-resolved magnetic anisotropy values, shown in Fig.~\ref{fig-layered}, are based on our first-principles calculations revealing that the interfacial magnetic anisotropy in Fe/MgO is mainly contributed from the first and second Fe monolayers~\cite{Hallal13}. The layer-resolved nearest-neighbor exchange constants are obtained as described in the previous subsection. The constrained Monte-Carlo approach, as described in Ref.~\cite{Asselin10}, is used to calculate the temperature-dependent saturation magnetization and anisotropy of the modeled system. We use $100,000$  Monte-Carlo equilibration steps followed by $100,000$ steps, with $10^{-16}$ s time-step, over which the thermodynamic averages are calculated to obtain the temperature-dependent system's properties. We note that the calculated temperature-dependent magnetization and anisotropy are normalized by their values at $T=10$ K so that to be closer to experiments done at very low temperatures where spin fluctuations are almost negligible.

\section{Choice of in-plane lattice parameter in the atomistic calculations}

As mentioned previously, our calculations are done in two steps. Our first-principles calculations start by optimizing the atomic structures through full relaxation of atomic positions and volume. In Fig.~\ref{fig-app3}, we show the variation of the relaxed in-plane lattice parameter of Fe/MgO as a function of Fe thickness. Here, one can clearly see how increasing the Fe thickness while keeping the MgO thickness fixed leads to a decrease in the lattice parameter getting closer to that of Fe $a=2.866$ \AA{}. Then, the calculated layer resolved magnetic moments, magnetic anisotropy energies, and exchange interactions are obtained using those relaxed structures. 

\begin{figure}[h]
	\centering
	\includegraphics[width=0.95\columnwidth]{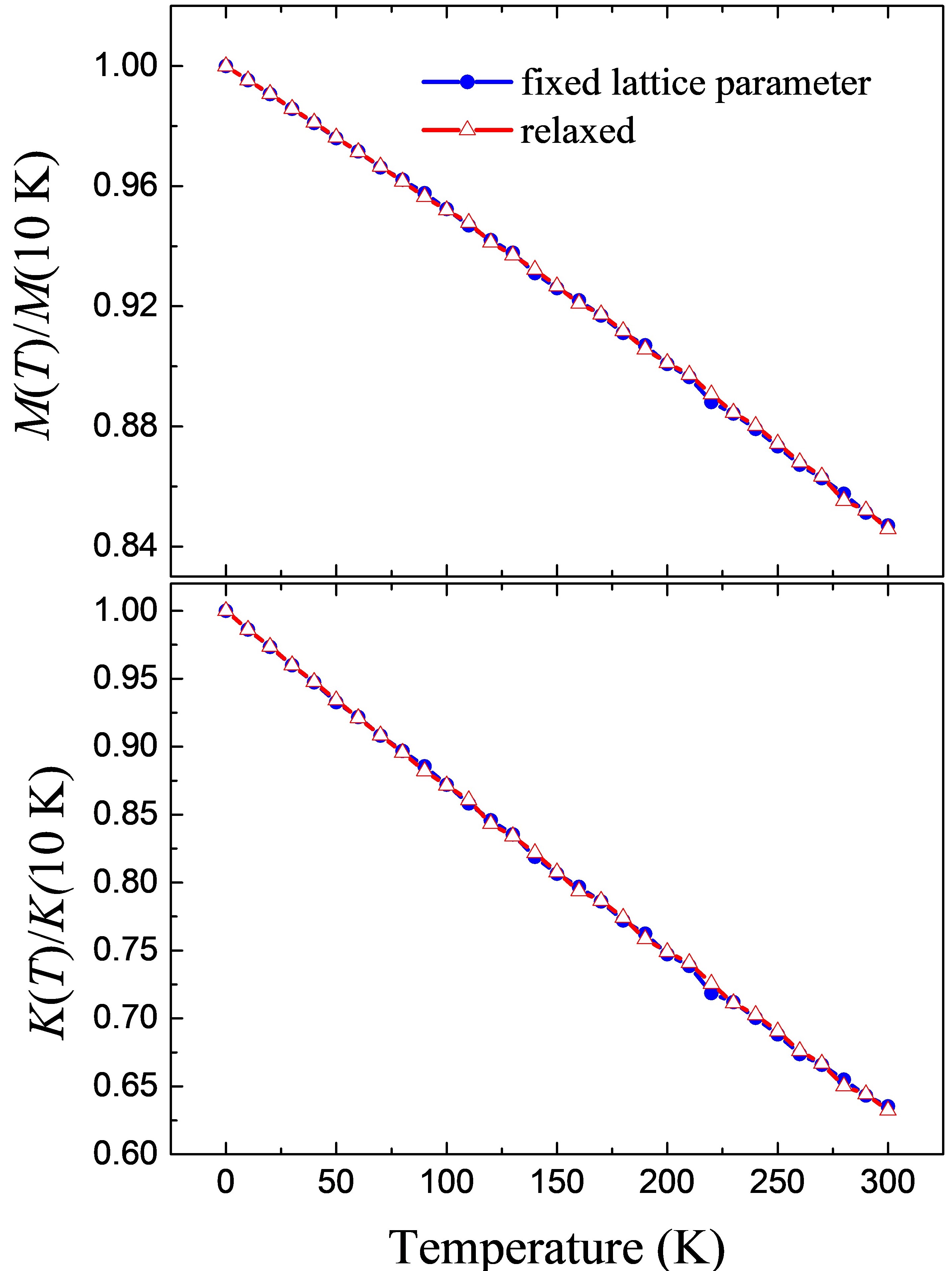}
	\caption{(Color online) The temperature-dependence of the normalized magnetization and anisotropy obtained from the atomistic calculations using either the ab initio optimized lattice constant or fixed to that of Fe.}
	\label{fig-app4}
\end{figure}

The atomistic calculations were parametrized with the layer resolved magnetic moments, anisotropy energies, and exchange interactions obtained from first-principles using optimized structures. Namely, we fixed the in-plane lattice parameter in the modeled cylindrical-shaped samples to $a=2.866$ \AA{}. In order to compare and verify that the effect of the lattice parameter is minor, we performed test calculations using the ab initio optimized lattice constants. For instance, Fig.~\ref{fig-app4} compares the results of those two calculations for 11 ML thick Fe samples. One can see that both the normalized saturation magnetization ($M$) and magnetic anisotropy ($K$) dependencies as a function of temperature are almost not affected by the choice of the lattice parameter. Consequently, the scaling exponent of their temperature dependence, which we focus on in the present study, is also not affected.

\section{Layer-resolved change of the magnetization as a function of temperature}

The decrease of the magnetization with temperature is indeed layer-dependent. In case of ideal interface, Fig.~\ref{fig-Fe2ML}(a) compares the layer-resolved saturation magnetization for the different layers. The magnetization of second iron layer (Fe2) decreases slower with temperature compared with that of the first (Fe1) and bulk region. This can be explained by the enhanced exchange at the interface, namely in nearest-neighbor pair interaction $J_{1,2}$ compared to bulk values accompanied by a reduced atomic coordination of the Fe1 layer and a full coordination of Fe2.

\begin{figure}[h]
	\centering
	\includegraphics[width=0.95\columnwidth]{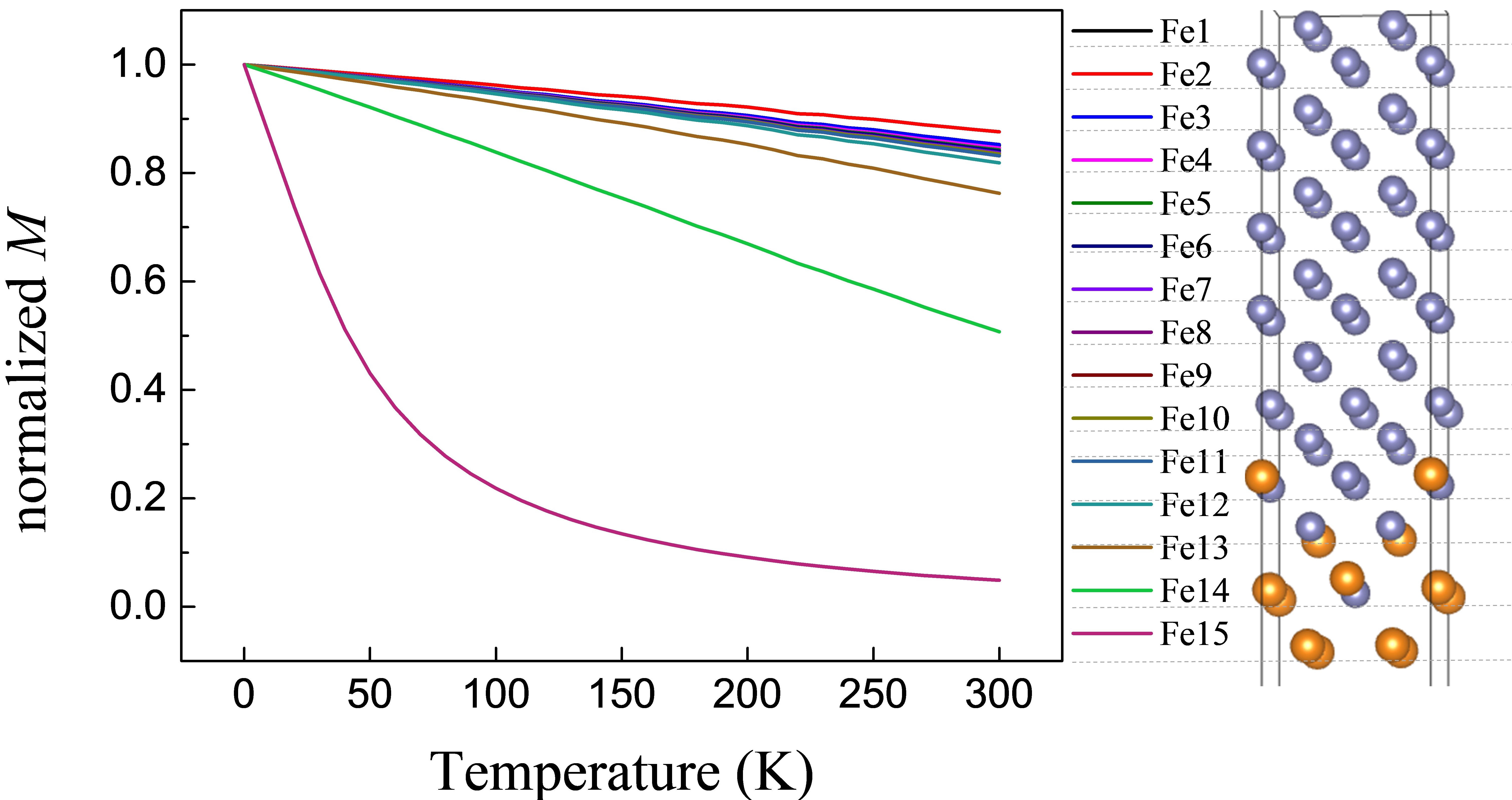}
	\caption{(Color online) The layer-resolved normalized magnetization as a function of temperature for a sample comprising a dead layer at its lower interface.}
	\label{fig-app5}
\end{figure}

In case of the presence of a dead layer, Fig.~\ref{fig-app5} shows layer-resolved variation of the normalized magnetization ($M$) as a function of temperature. The previous explanation still holds for the upper interface which is ideal and we can observe almost identical trend for all the bulk layers. However, a faster decrease of $M$ is found for the lower interface. This is due to the reduced exchange caused by the Ta diffusion, namely in Fe13-Fe14-Fe15, that can be also depicted in Fig.~\ref{fig-dead-layer-exchange}.




\bibliography{NC10238}

\end{document}